\documentstyle[12pt]{article}
\newcommand{\be}{\begin{equation}}
\newcommand{\ee}{\end{equation}}
\newcommand{\bea}{\begin{array}}
\newcommand{\ea}{\end{array}}
\newcommand{\beqa}{\begin{eqnarray}}
\newcommand{\eeqa}{\end{eqnarray}}
\newcommand{\bean}{\begin{eqnarray*}}
\newcommand{\eean}{\end{eqnarray*}}
\newcommand{\eqn}[1]{(\ref{#1})}

\def\ru1{\rule[-0.4truecm]{0mm}{1truecm}}

\def\lrw{\longrightarrow}
\def\lu{SU(3)_c\otimes SU(2)_L\otimes SU(2)_R\otimes U(1)_{B-L}}

\def\SM{G_{SM}}
\def\ul{SU(3)_c\otimes U(1)_Q}
\def\tiz{SU(4)_{PS}\otimes SU(2)_L\otimes SU(2)_R}

\def\up#1{\leavevmode \raise.16ex\hbox{#1}}

\newcommand{\jou}[4]{{\sl #1 }{\bf #2} \up(19#3\up) #4}

\begin{document}

\title{\hfill $\mbox{\small{
$\stackrel{\rm\textstyle DSF/53-97}
{\rm\textstyle hep-ph/9711306\quad\quad}$}}$ \\[1truecm]
Limits on neutrino masses in SO(10) GUT's\footnote{Talk given by L. Rosa at
the International School on Nuclear Physics; 19th Course: Neutrinos in
Astro, Particle and Nuclear Physics, Erice, September 16-24, 1997.}} 
\author{O. Pisanti$^\dagger$ and L. Rosa$^\dagger$}
\date{$~$}

\maketitle

\begin{center}
\begin{tabular}{l}
$^\dagger$  Dipartimento di Scienze Fisiche, Universit\`a di Napoli, \\
~~Mostra d'Oltremare, Pad.19, I-80125 Napoli, Italy; \\
~~INFN, Sezione di Napoli, I-80125 Napoli, Italy.\\ \\
\small\tt e-mail: pisanti@na.infn.it  \\
\small\tt e-mail: rosa@na.infn.it 
\end{tabular}
\end{center}

\begin{abstract}
Studying Renormalization Group Equations for the four typical SO(10)
spontaneous symmetry breaking patterns we find the existence of lower
limits on neutrino masses. 
\end{abstract}

SO(10) \cite{geor} models can be characterized by the intermediate symmetry
group $(G')$ appearing when SO(10) breaks down to $SU(3)\otimes
SU(2)\otimes U(1)~(\equiv G_{SM})$, 
{
\small
\be
SO(10)\buildrel M_X\over\lrw G'\buildrel M_R\over\lrw\SM\buildrel M_Z\over
\lrw\ul. \label{eq:ssb}
\ee
}
$M_X$ and $M_R$ are connected to the proton lifetime \cite{bumi} and
Majorana neutrino masses \cite{gera} respectively, 
\be
\tau_{p\rightarrow e^+\pi^0}= 1.1\div1.4~10^{32} 
\left( \frac{M_X}{10^{15}~GeV} \right)^4~years,
~~~~~m_{\nu_i} = \left( \frac{m_\tau}{m_b} \right)^2 \frac{m_{u_i}^2}{M_R} 
\frac{g_{2R}(M_R)}{y_i(M_R)}, \label{eq:limits}
\ee
where $u_i$ is the up quark of the i-th generation, $g_{2R}$ is the
$SU(2)_R$ coupling and $y_i$ the Yukawa couplings of the $i$-th fermion
generation to the Higgs responsible for the symmetry breaking at $M_R$. 

Depending on the representations used to classify the Higgses and on the
direction of minimum chosen by the Higgs potential, we obtain the following
four patterns of symmetry breaking \cite{acam,lemo}, which represent the
four simplest and realistic minimal SO(10) models: 

{
\footnotesize
\[
  \bea{ccl}
    SO(10) &  Higgses &  ~~~~~~~~~~~~~~~~G'                 \\ 
    (i)  &   \buildrel \sigma\over\lrw   & G_{422D}\equiv \tiz\times D   \\ 
    (ii)  &  \buildrel \Phi_1\over\lrw  & G_{3221D}\equiv \lu\times D    \\ 
    (iii)  &  \buildrel \Phi_2\over\lrw  & G_{422}\equiv \tiz     \\ 
    (iv)   &  \buildrel \Phi_3\over\lrw  & G_{3221}\equiv \lu,
  \ea
\]
}

\noindent($\sigma$ is the 54 second rank symmetric representation while 
$\Phi$ is the 210 fourth rank antisymmetric representation of SO(10),
$\Phi_1=\frac{1}{\sqrt{3}}(\Phi_{1234}+\Phi_{3456}+\Phi_{5612} )$,
$\Phi_2=\Phi_{78910}$ and ${\Phi_3}= cos\theta{\Phi_1} + sin
\theta{\Phi_2}$). 

$M_X$ and $M_R$ are found by solving the following Renormalization Group
Equations (RGE) (at first loop)\footnote{We use two 10-dimensional
representations so to avoid the unwanted relation $m_t=m_b$. }: 
\beqa
\frac{\pi}{11 \alpha(M_Z)} \left( \sin^2\theta_W(M_Z)-
\frac{\alpha}{\alpha_S}(M_Z)  \right) & = & 
\frac{12}{11}\ln\frac{M_R}{M_Z}+C_1\ln\frac{M_X}{M_R}, \\
\frac{\pi}{6 \alpha(M_Z)} \left(\frac{3}{2}-3 \sin^2\theta_W(M_Z)-
\frac{\alpha}{\alpha_S}(M_Z)  \right)  & = & \ln\frac{M_R}{M_Z}+
C_2\ln\frac{M_X}{M_R},
\eeqa
where $C_1$ and $C_2$ parametrize the dependence of RGE on $G'$ and on the
Higgses \cite{acam}. 

\begin{table}[t]
\begin{center}
TABLE \ref{t:mass} \\ \medskip
\begin{tabular}{||c||c|c||c||c|c|}
\hline\ru1
$G'$ & $M_X/10^{15}\,GeV$ & $M_R/10^{11}\,GeV$ &
$G'$ & $M_X/10^{15}\,GeV$ & $M_R/10^{11}\,GeV$ \\
\hline\ru1
$G_{422D}$ & 0.6 & 420 & $G_{422}$ & 30 & 1.1 \\
\hline\ru1
$G_{3221D} $ & 1.9 & 0.78 & $G_{3221}$ & 15 & 0.26 \\
\hline  
\end{tabular}
\end{center}
\caption{Values of $M_X$ and $M_R$,
$\sin^2{\theta_w}=0.2315\pm0.0002;~\alpha_s=0.118\pm0.003;~
\frac{1}{\alpha}=127.9\pm0.09$ \protect\cite{pdb}} 
\label{t:mass}
\end{table}

A thorough study of these RGE has been carried in \cite{acam}, making it
possible to show that in the four cases an upper limit on $M_R$ exists,
which is, trough \eqn{eq:limits}, a lower limit on $m_{\nu_i}$. In table
\ref{t:mass} we report the highest values of $M_R$ obtained in the four
cases together with the corresponding value of $M_X$. We see that model
$(i)$ is ruled out by the limit on proton decay \cite{pdb}
$M_X\geq1.6~10^{15}~GeV$. Model $(iv)$ gives ($g_{2R}\sim y_i$)
$m_{\nu_\tau}\geq200~eV$, too high with respect to the Cowsik Mc Clelland
limit $\sum_i{ m_{\nu_i}}\leq100~eV$ \cite{cocl}. Model $(ii)$ can
eventually be ruled out by the next running of SuperKamiokande \cite{pdb}.
Model $(iii)$ gives $m_{\nu_\tau}\geq50~eV$, the order of magnitude
required for the hot solution to the dark matter problem, while
$m_{\nu_\mu}\sim10^{-3}~eV$ is of the order required by the MSW solution
\cite{miwo} to solar-neutrino problem.

\end{document}